\documentclass[ aps, showpacs, showkeys, nofootinbib, floatfix, superscriptaddress]{revtex4}

\usepackage{amsfonts}

\usepackage{amssymb}
\usepackage{amsmath}
\usepackage{graphicx}

\begin{document}

\title{Modified $f(G)$ gravity
models with curvature-matter coupling}

 \author{Yue-Yue Zhao}
 \email{zhaoyueyue198737@163.com}
 \affiliation{Department of Physics, Liaoning Normal University, Dalian 116029, P. R. China}
\author{Ya-Bo Wu}
 \email{ybwu61@163.com}
 \affiliation{Department of Physics, Liaoning Normal University, Dalian 116029, P. R. China}
\author{Jian-Bo Lu}
 \affiliation{Department of Physics, Liaoning Normal University, Dalian 116029, P. R. China}
\author{Zhuo Zhang}
 \affiliation{Department of Physics, Liaoning Normal University, Dalian 116029, P. R. China}
\author{Wei-Li Han}
 \affiliation{Department of Physics, Liaoning Normal University, Dalian 116029, P. R. China}
\author{Liang-Liang Lin}
 \affiliation{Department of Physics, Liaoning Normal University, Dalian 116029, P. R. China}

\begin{abstract}

A modified f(G) gravity model with coupling between matter and
geometry is proposed, which is described by the product of the
Lagrange density of the matter and an arbitrary function of the
Gauss-Bonnet term. The field equations and the equations of motion
corresponding to this model show the non-conservation of the
energy-momentum tensor, the presence of an extra-force acting on
test particles and the non-geodesic motion. Moreover, the energy
conditions and the stability criterion at de Sitter point in the
modified f(G) gravity models with curvature-matter coupling are
derived, which can degenerate to the well-known energy conditions in
general relativity. Furthermore, in order to get some insight on the
meaning of these energy conditions, we apply them to the specific
models of f(G) gravity and the corresponding constraints on the
models are given. In addition, the conditions and the candidate for
late-time cosmic accelerated expansion in the modified f(G) gravity
are studied by means of conditions of power-law expansion and the
equation of state of matter less than $-\frac{1}{3}$.

\end{abstract}



\maketitle

\section{Introduction}

~~~~According to recent observational data sets\cite{1,2}, our
current universe is flat and undergoing a phase of the accelerated
expansion which started about five billion years ago. In principle,
this phenomenon can be explained by either dark energy (see, for
instance, Ref.\cite{3} for reviews), in which the reason of this
phenomenon is due to an exotic component with large negative
pressure, or modified theories of gravity\cite{4}. Unfortunately, up
to now a satisfactory answer to the question that what dark energy
is and where it came from has not yet to be obtained. Alternative to
dark energy, modified theories of gravity is extremely attractive,
such as f(R) gravity (see, for instance, Ref.\cite{5} for reviews),
here f(R) is an arbitrary function of the Ricci scalar R. Cosmic
acceleration can be explained by f(R) gravity\cite{6}, and the
conditions of viable cosmological models have been derived in
\cite{7}.

A general model of $f(R)$ gravity has been proposed in Ref.\cite{8},
which contains a non-minimal coupling between geometry and matter.
This coupling term can be considered as a gravitational source to
explain the current acceleration of the universe. As a result of the
coupling the motion of the massive particles is non-geodesic, and an
extra force, orthogonal to the four-velocity, arises. Different
forms for the matter Lagrangian density $L_{m}$, and the resulting
extra-force, were considered in \cite{9}, and it was shown that more
natural forms for $L_{m}$ do not imply the vanishing of the
extra-force. The implications of the non-minimal coupling on the
stellar equilibrium were investigated in \cite{10}, where
constraints on the coupling were also obtained. An inequality which
expresses a necessary and sufficient condition to avoid the
Dolgov-Kawasaki instability for the model was derived in \cite{11}.
However, a more general model, in which the coupling style is
arbitrary and the Lagrangian density of matter only appears in
coupling term, has been proposed in Ref.\cite{12}, i.e., the
so-called the generalized f(R) gravity with arbitrary coupling
between matter and geometry. In this class of models the
energy-momentum tensor of the matter is generally not conserved, and
the matter-geometry coupling can induce a supplementary acceleration
of the test particles. Moreover, the energy conditions and  the
Dolgov-Kawasaki criterion for the model have been derived in
Ref.\cite{13}, which are quite general and can degenerate to the
well-known energy conditions in GR and f(R) gravity with non-minimal
coupling and non-coupling as special cases.

Another interesting alternative modified theory of gravity is the
modified Gauss-Bonnet gravity, i.e., f(G) gravity, where f(G) is a
general function of the Gauss-Bonnet (GB) term\cite{14,15}. At
present specific models of f(G) gravity have been proposed to
account for the late-time cosmic acceleration\cite{15,16}, and the
respective constraints on the parameters of the models have also
analyzed in Ref.\cite{16}. For more general forms of f(G), the most
crucial condition to be satisfied is $d^{2}f/dG^{2}>0$, which is
required to ensure the stability of a late-time de Sitter solution
as well as the existence of standard radiation/matter dominated
epochs\cite{17}, and solar system constraints on f(G) gravity models
have been studied in Ref.\cite{18}. Recently the energy conditions
in f(G) gravity have been also discussed\cite{19}, but they are only
adapted to $f(G)$ gravity without coupling between matter and
geometry. Hence, in this paper the $f(G)$ gravity models with
curvature-matter coupling will be proposed, and some relevant
issues, such as the energy conditions, the stability criterion and
the conditions for late-time cosmic accelerated expansion, will be
studied.

This paper is organized as follows. In Section 2, the $f(G)$ gravity
models with curvature-matter coupling are proposed, which is here
called the modified $f(G)$ gravity. And some fundamental elements of
the modified $f(G)$ gravity are given. In Section 3, the well-known
energy conditions, namely, the strong energy condition (SEC), the
null energy condition (NEC), the weak energy condition (WEC) and the
dominant energy condition (DEC), will be derived. Concretely, the
two models are applied to the weak energy condition in order to
understand the meaning of these energy conditions. Furthermore, we
will study stability criterion at the de Sitter point, by which the
parameters in the specific model in the modified f(G) gravity can be
constrained in Section 4. In addition, by using the conditions of
power-law accelerated expansion and the equation of state of matter
less than $-\frac{1}{3}$, the conditions for late-time cosmic
accelerated expansion in the modified f(G) gravity are discussed in
Section 5. Conclusions and discussions on our work are given in the
last section.

\section{Field equations in the modified f(G) gravity}

~~~~As we know, the modified Einstein-Hilbert action\cite{20} is,
\begin{equation} \label{1}
S=\int d^4x\sqrt{-g}[\frac{R+f(G)}{2\kappa}  +L_m],
\end{equation}
in which $\kappa=8\pi G_{N}$, $G_{N}$ is the gravitational constant,
$R=R(g_{\mu\nu})$ is the Ricci scalar, and $L_{m}$ is the Lagrangian
density of matter. The Gauss-Bonnet invariant is defined as $G\equiv
R^{2}-4R_{\mu\nu}R^{\mu\nu}+R_{\mu\nu\xi\sigma}R^{\mu\nu\xi\sigma}$
($R_{\mu\nu}$ and $R_{\mu\nu\xi\sigma}$ are the Ricci tensor and the
Riemann tensor, respectively).

Below, we consider f(G) gravity with curvature-matter coupling,
which is here called the modified f(G) gravity. The Lagrangian
density of matter only appears in coupling term and the action is
given by
\begin{equation} \label{2}
S=\int d^4x\sqrt{-g}\{\frac{R}{2}+[1+\lambda f(G)]L_m\},
\end{equation}
where we have chosen $\kappa=8\pi G_{N}=1$, which will be adopted
hereafter. Varying the action (\ref{2}) with respect to the metric
$g^{\mu\nu}$ yields the field equations:
\begin{equation}\label{3}
\begin{array}{rcl}
R_{\mu\nu}-\frac{1}{2}g_{\mu\nu}R&=& T_{\mu\nu}+2\lambda
L_{m}(-2FRR_{\mu\nu}+4FR^{\xi}_{\mu}R_{\nu\xi}
-2FR_{\mu\xi\sigma\varsigma}R^{\xi\sigma\varsigma}_{\nu}-4FR_{\mu\xi\sigma\nu}R^{\xi\sigma}\\
&&+2R\nabla_{\mu}\nabla_{\nu}F-2Rg_{\mu\nu}\nabla^{2}F-4R^{\xi}_{\mu}\nabla_{\nu}\nabla_{\xi}F
-4R^{\xi}_{\nu}\nabla_{\mu}\nabla_{\xi}F
\\&&+4R_{\mu\nu}\nabla^{2}F
+4g_{\mu\nu}R^{\xi\sigma}\nabla_{\xi}\nabla_{\sigma}F-4R_{\mu\xi\nu\sigma}\nabla^{\xi}\nabla^{\sigma}F),
\end{array}
\end{equation}
where $F=F(G)\equiv\partial f(G)/\partial G$, and $T_{\mu\nu}$ is
the energy-momentum tensor of matter, which is defined as:
\begin{equation}\label{4}
T_{\mu\nu}=-\frac{2}{\sqrt{-g}}\cdot \frac{\delta
(\sqrt{-g}L_{m})}{\delta g^{\mu\nu}}.
\end{equation}

By assuming that the Lagrangian density $L_{m}$ of the matter
depends only on the metric  tensor components, and not on its
derivatives, we obtain $T_{\mu\nu}=L_{m}g_{\mu\nu}-2\partial
L_{m}/\partial g^{\mu\nu}$. By taking the covariant derivative of
Eq.(\ref{3}), with the use of the Bianchi identities,
$\nabla^{\mu}G_{\mu\nu}=0$ ( here $G_{\mu\nu}$ is the Einstein
tensor ), we can obtain the following relation:
\begin{equation}\label{5}
\begin{array}{rcl}
\nabla^{\mu}T_{\mu\nu}&=&4\lambda
L_{m}[\nabla^{\mu}F(RR_{\mu\nu}-2R^{\xi}_{\mu}R_{\nu\xi}+R_{\mu\xi\sigma\varsigma}R^{\xi\sigma\varsigma}_{\nu}
+2g^{\alpha\xi}g^{\beta\sigma}R_{\mu\alpha\nu\beta}R_{\xi\sigma})\\&&+F(R_{\mu\nu}\nabla^{\mu}R+\frac{1}{2}g_{\mu\nu}R\nabla^{\mu}R
-2R_{\nu\xi}\nabla^{\mu}R^{\xi}_{\mu}-2R^{\xi}_{\mu}\nabla^{\mu}R_{\nu\xi}+R^{\xi\sigma\varsigma}_{\nu}\nabla^{\mu}R_{\mu\xi\sigma\varsigma}
\\&&+R_{\mu\xi\sigma\varsigma}\nabla^{\mu}R^{\xi\sigma\varsigma}_{\nu}+2g^{\alpha\xi}g^{\beta\sigma}R_{\xi\sigma}\nabla^{\mu}R_{\mu\alpha\nu\beta}
+2g^{\alpha\xi}g^{\beta\sigma}R_{\mu\alpha\nu\beta}\nabla^{\mu}R_{\xi\sigma})+R\nabla_{\nu}\Box
F\\&&-R\Box\nabla_{\nu}F-\nabla_{\mu}\nabla_{\nu}F\nabla^{\mu}R-2R_{\mu\nu}\nabla^{\mu}\Box
F+2R^{\xi}_{\mu}\nabla^{\mu}\nabla_{\nu}\nabla_{\xi}F+2\nabla_{\nu}\nabla_{\xi}F\nabla^{\mu}R^{\xi}_{\mu}\\&&+2R^{\xi}_{\nu}\Box\nabla_{\xi}F
+2\nabla_{\mu}\nabla_{\xi}F\nabla^{\mu}R^{\xi}_{\nu}-2g_{\mu\nu}R^{\xi\sigma}\nabla^{\mu}\nabla_{\xi}\nabla_{\sigma}F
-2g_{\mu\nu}\nabla_{\xi}\nabla_{\sigma}F\nabla^{\mu}R^{\xi\sigma}\\&&+2g^{\alpha\xi}g^{\beta\sigma}R_{\mu\alpha\nu\beta}\nabla^{\mu}\nabla_{\xi}\nabla_{\sigma}F
+2g^{\alpha\xi}g^{\beta\sigma}\nabla_{\xi}\nabla_{\sigma}F\nabla^{\mu}R_{\mu\alpha\nu\beta}],
\end{array}
\end{equation}
from which we see that the conservation of the energy-momentum
tensor of matter is violated due to the coupling between matter and
geometry.

For the perfect fluid, the energy-momentum tensor is
\begin{equation}\label{6}
T_{\mu\nu}=(\rho+p)u_{\mu}u_{\nu}+pg_{\mu\nu},
\end{equation}
where $u_{\mu}$ is the four-velocity, and satisfies the conditions
$u_{\mu}u^{\mu}=1$ and $u^{\mu}u_{\mu;\nu}=0$\cite{8}. Thus, the
covariant derivative of Eq.(\ref{6}) can be given as
\begin{equation}\label{7}
\nabla^{\mu}T_{\mu\nu}=(\rho+p)g_{\mu\lambda}u^{\nu}\nabla_{\nu}u^{\mu}-\nabla_{\nu}p~(\delta^{\nu}_{\lambda}-u^{\nu}u_{\lambda}).
\end{equation}
By imposing the condition of the conservation of the matter current,
$\nabla_{\nu}(\rho u^{\nu})=0$, and with the use of the identity
$u^{\nu}\nabla_{\nu}u^{\mu}=\frac{d^{2}x^{\mu}}{ds^{2}}+\Gamma^{\mu}_{\nu\lambda}u^{\nu}u^{\lambda}$\cite{12},
we have the equation of motion of a test particle in the model as
\begin{equation}\label{8}
\frac{Du^{\mu}}{ds}\equiv
\frac{du^{\mu}}{ds}+\Gamma^{\mu}_{\nu\lambda}u^{\nu}u^{\lambda}=\frac{d^{2}x^{\mu}}{ds^{2}}+\Gamma^{\mu}_{\nu\lambda}u^{\nu}u^{\lambda}=f^{\mu},
\end{equation}
where the extra-force $f^{\mu}$ has the following expression
\begin{equation}\label{9}
f^{\mu}=\frac{1}{\rho+p}[\nabla^{\mu}T_{\mu\nu}g^{\mu\lambda}+\nabla_{\nu}p~(g^{\mu\nu}-u^{\nu}u^{\mu})].
\end{equation}
By substituting the relation (\ref{5}) into Eq.(\ref{9}), we can
point that due to the presence of the coupling between matter and
geometry, the motion of the massive particles is non-geodesic, and
the extra-force $f^{\mu}$ is not orthogonal to the four-velocity
$u_{\mu}$, that is, there exists an angle between the extra-force
$f^{\mu}$ and the four-velocity $u_{\mu}$, and
$f^{\mu}u_{\mu}\neq0$.

\section{Energy conditions in the modified f(G) gravity }

\subsection{The Raychaudhuri Equation}

~~~The energy conditions arise when one refers to the Raychaudhuri
equation for the expansion\cite{21}.  Under these energy conditions,
one allows not only to establish gravity which remains attractive,
but also to keep the demands that the energy density is positive and
cannot flow faster than light. Below, following Ref.\cite{13} we
simply review the Raychaudhuri equation which is the physical origin
of the null energy condition(NEC) and the strong energy
condition(SEC)\cite{22}.

In the case of a congruence of timelike geodesics defined by the
vector field $u^{\mu}$, the Raychaudhuri equation is given by
\begin{equation}\label{10}
\frac{d\theta}{d\tau}=-\frac{1}{3}\theta^{2}-\sigma_{\mu\nu}\sigma^{\mu\nu}
+\omega_{\mu\nu}\omega^{\mu\nu}-R_{\mu\nu}u^{\mu}u^{\nu},
\end{equation}
where $R_{\mu\nu},\theta,\sigma_{\mu\nu}$ and $\omega_{\mu\nu}$ are
the Ricci tensor, the expansion parameter, the shear and the
rotation associated with the congruence, respectively. While in the
case of a congruence of null geodesics defined by the vector field
$k^{\mu}$, the Raychaudhuri equation is given by
\begin{equation}\label{11}
\frac{d\theta}{d\tau}=-\frac{1}{2}\theta^{2}-\sigma_{\mu\nu}\sigma^{\mu\nu}+\omega_{\mu\nu}\omega^{\mu\nu}
-R_{\mu\nu}k^{\mu}k^{\nu}.
\end{equation}

From above expressions, it is clear that the Raychaudhuri equation
is a purely geometric statement and independent of the gravity
theory. In order to constrain the energy-momentum tensor by the
Raychaudhuri equation, one can use the Ricci tensor from the field
equations of gravity to make a connection. Namely, through the
combination of the field equations of gravity and the Raychaudhuri
equation, one can obtain physical conditions for the energy-momentum
tensor. Since $\sigma^{2} \equiv \sigma_{\mu\nu}\sigma^{\mu\nu} \geq
0$ (the shear is a spatial tensor) and $\omega_{\mu\nu}=0$
(hypersurface orthogonal congruence), from Eqs. (\ref{10}) and
(\ref{11}), the conditions for gravity to remain attractive
($\frac{d\theta}{d\tau}<0$) are
\begin{equation}\label{12}
R_{\mu\nu}u^{\mu}u^{\nu}\geq 0 ~~~~SEC,
\end{equation}
\begin{equation}\label{13}
R_{\mu\nu}k^{\mu}k^{\nu}\geq 0 ~~~~NEC.
\end{equation}

Thus by means of the relationship (\ref{12}) and Einstein's equation
($R_{\mu\nu}-\frac{1}{2}Rg_{\mu\nu}=T_{\mu\nu}$), one obtains
\begin{equation}\label{14}
R_{\mu\nu}u^{\mu}u^{\nu}=(T_{\mu\nu}-\frac{T}{2}g_{\mu\nu})u^{\mu}u^{\nu}
\geq 0.
\end{equation}
If one considers a perfect fluid with energy density $\rho$ and
pressure p,
\begin{equation}\label{15}
T_{\mu\nu}=(\rho+p)u_{\mu}u_{\nu}+pg_{\mu\nu}
\end{equation}
the relationship (\ref{14}) turns into the well-known SEC in general
relativity, i.e.,
\begin{equation}\label{16}
\rho+3p \geq 0.
\end{equation}
Similarly, by using the relationship (\ref{13}) and Einstein's
equation, one has
\begin{equation}\label{17}
T_{\mu\nu}k^{\mu}k^{\nu}\geq 0.
\end{equation}
Thus, by considering Eq.(\ref{15}), the familiar NEC in general
relativity can be reproduced as:
\begin{equation}\label{18}
\rho+p \geq 0.
\end{equation}

\subsection{Energy conditions}

~~~The Einstein tensor resulting from the field equation (\ref{3})
can be written
\begin{equation}\label{19}
G_{\mu\nu}\equiv R_{\mu\nu}-\frac{1}{2}g_{\mu\nu}R=T^{eff}_{\mu\nu},
\end{equation}
where
\begin{equation}\label{20}
\begin{array}{rcl}
T^{eff}_{\mu\nu}&=&T_{\mu\nu}+2\lambda
L_{m}(-2FRR_{\mu\nu}+4FR^{\xi}_{\mu}R_{\nu\xi}
-2FR_{\mu\xi\sigma\varsigma}R^{\xi\sigma\varsigma}_{\nu}-4FR_{\mu\xi\sigma\nu}R^{\xi\sigma}
\\&&+2R\nabla_{\mu}\nabla_{\nu}F-2Rg_{\mu\nu}\nabla^{2}F
-4R^{\xi}_{\mu}\nabla_{\nu}\nabla_{\xi}F-4R^{\xi}_{\nu}\nabla_{\mu}\nabla_{\xi}F
\\&&+4R_{\mu\nu}\nabla^{2}F
+4g_{\mu\nu}R^{\xi\sigma}\nabla_{\xi}\nabla_{\sigma}F-4R_{\mu\xi\nu\sigma}\nabla^{\xi}\nabla^{\sigma}F).
\end{array}
\end{equation}
Contracting the above equation, we have
\begin{equation}\label{21}
T^{eff}=T-4\lambda L_{m}(FG+R\Box
F-2R_{\mu\nu}\nabla^{\mu}\nabla^{\nu}F).
\end{equation}
Thus, we can write $R_{\mu\nu}$ in terms of an effective
stress-energy tensor and its trace, i.e. ,
\begin{equation}\label{22}
R_{\mu\nu}=T^{eff}_{\mu\nu}-\frac{1}{2}g_{\mu\nu}T^{eff}.
\end{equation}
By using the relationship (\ref{12}) and Eq.(\ref{22}), the SEC can
be given as:
\begin{equation}\label{23}
T_{\mu\nu}^{eff}u^{\mu}u^{\nu}-\frac{1}{2}T^{eff} \geq 0.
\end{equation}
By using the Eqs.(\ref{20}) and (\ref{21}), the SEC in Eq.(\ref{23})
can be expressed as
\begin{equation}\label{24}
\rho+3p+8\lambda L_{m}FG-96\lambda L_{m}H^{3}\dot{f_{,G}}+4\lambda
L_{m}R\Box F-8\lambda L_{m}R_{\mu\nu}\nabla^{\mu}\nabla^{\nu}F \geq
0,
\end{equation}
where $H=\dot{a}(t)/a(t)$ is the Hubble parameter. The NEC in the
modified f(G) gravity can be expressed as:
\begin{equation}\label{25}
T_{\mu\nu}^{eff}k^{\mu}k^{\nu} \geq 0.
\end{equation}
By the same method as the SEC, the NEC in Eq.(\ref{25}) can be
changed into
\begin{equation}\label{26}
\rho+p+4\lambda L_{m}FG-64\lambda
L_{m}H^{3}\dot{f_{,G}}+\frac{4}{3}\lambda L_{m}R\Box F -
\frac{8}{3}\lambda L_{m}R_{\mu\nu}\nabla^{\mu}\nabla^{\nu}F \geq 0.
\end{equation}
Furthermore, by means of Eqs.(\ref{20}) and (\ref{22}), the
effective energy density and the effective pressure can be derived
as follows:
\begin{equation}\label{27}
\rho^{eff}=2\lambda L_{m}(FG-24H^{3}\dot{f_{,G}})+\rho,
\end{equation}
\begin{equation}\label{28}
p^{eff}=2\lambda L_{m}FG-16\lambda
L_{m}H^{3}\dot{f_{,G}}+\frac{4}{3}\lambda L_{m}R\Box
F-\frac{8}{3}\lambda L_{m}R_{\mu\nu}\nabla^{\mu}\nabla^{\nu}F+p.
\end{equation}

Note that the above expressions of the SEC (\ref{24}) and the NEC
(\ref{26}) are directly derived from Raychaudhuri equation. However,
equivalent results can obtained by the transformations $\rho
\rightarrow \rho^{eff}$ and $p \rightarrow p^{eff}$ into
$\rho+3p\geq0$ and $\rho+p\geq0$. Thus by extending this approach to
$\rho-p\geq0$ and $\rho\geq0$, the corresponding DEC and the WEC in
the modified f(G) gravity can be respectively given as:
\begin{equation}\label{29}
\rho-p-32\lambda L_{m}H^{3}\dot{f_{,G}}-\frac{4}{3}\lambda
L_{m}R\Box F+\frac{8}{3}\lambda
L_{m}R_{\mu\nu}\nabla^{\mu}\nabla^{\nu}F\geq0,
\end{equation}
\begin{equation}\label{30}
\rho+2\lambda L_{m}FG-48H^{3}\dot{f_{,G}}\geq0.
\end{equation}
It is worth stressing that when taking $f(G)=0$, the energy
conditions in general relativity can be reproduced.

Furthermore, by defining the deceleration, jerk, and snap parameters
as\cite{23}
\begin{equation}\label{31}
q=-\frac{1}{H^{2}}\cdot\frac{\ddot{a}}{a},
~j=\frac{1}{H^{3}}\cdot\frac{\dot{\ddot{a}}}{a},
~s=\frac{1}{H^{4}}\cdot\frac{\ddot{\ddot{a}}}{a},
\end{equation}
we have
\begin{equation}\label{32}
\dot{H}=-H^{2}(1+q), \ddot{H}=H^{3}(j+3q+2),
\dot{\ddot{H}}=H^{4}(s-2j-5q-3),
\end{equation}
and the GB term is given by
\begin{equation}\label{33}
G=24H^{2}(H^{2}+\dot{H}).
\end{equation}
In addition, we consider $L_{m}=-\rho$, we can rewrite Eqs.
(\ref{27}) and (\ref{28}) as follows:
\begin{equation}\label{34}
\rho^{eff}=\rho+48\lambda H^{4}\rho[qf'+24H^{4}(2q^{2}+3q+j)f''],
\end{equation}
\begin{equation}\label{35}
p^{eff}=p+48\lambda
H^{4}\rho[qf'+4H^{4}(-2q^{3}-8q^{2}+q-2j-6qj+s+3)f''+96H^{8}(2q^{2}+3q+j)^{2}f'''].
\end{equation}

Hence, the energy conditions, i.e., the SEC, NEC, DEC and WEC can be
rewritten as:
\begin{eqnarray}
\rho+3p+48\lambda H^{4}\rho[q f'+24H^{4}(2q^{2}+3q+j)f'']+144\lambda
H^{4}\rho[q
f'+4H^{4}(-2q^{3}\nonumber\\-8q^{2}+q-2j-6qj+s+3)f''+96H^{8}(2q^{2}+3q+j)^{2}f''']\geq0,
\end{eqnarray}
\begin{eqnarray}
\rho+p+48\lambda H^{4}\rho[q f'+24H^{4}(2q^{2}+3q+j)f''] +48\lambda
H^{4}\rho[q f'+4H^{4}(-2q^{3}\nonumber\\-8q^{2}+q-2j-6qj+s+3)f''
+96H^{8}(2q^{2}+3q+j)^{2}f''']\geq0,
\end{eqnarray}
\begin{eqnarray}
\rho-p+48\lambda H^{4}\rho[qf'+24H^{4}(2q^{2}+3q+j)f'']-48\lambda
H^{4}\rho[q f'+4H^{4}(-2q^{3}\nonumber\\-8q^{2}+q-2j-6qj+s+3)f''
+96H^{8}(2q^{2}+3q+j)^{2}f''']\geq0,
\end{eqnarray}
\begin{equation}\label{39}
\rho+48\lambda H^{4}\rho[qf'+24H^{4}(2q^{2}+3q+j)f'']\geq0.
\end{equation}

It is worth stressing that when taking $L_{m}=p$, Eqs.(\ref{27}) and
(\ref{28}) change into
\begin{equation}\label{40}
\rho^{eff}=\rho-48\lambda H^{4}p[qf'+24H^{4}(2q^{2}+3q+j)f''],
\end{equation}
\begin{equation}\label{41}
p^{eff}=p-48\lambda
H^{4}p[qf'+4H^{4}(-2q^{3}-8q^{2}+q-2j-6qj+s+3)f''+96H^{8}(2q^{2}+3q+j)^{2}f'''].
\end{equation}
And the corresponding energy conditions are as follows:
\begin{eqnarray}
\rho+3p-48\lambda H^{4}p[q f'+24H^{4}(2q^{2}+3q+j)f'']-144\lambda
H^{4}p[q
f'+4H^{4}(-2q^{3}\nonumber\\-8q^{2}+q-2j-6qj+s+3)f''+96H^{8}(2q^{2}+3q+j)^{2}f''']\geq0,
\end{eqnarray}
\begin{eqnarray}
\rho+p-48\lambda H^{4}p[q f'+24H^{4}(2q^{2}+3q+j)f'']-48\lambda
H^{4}p[q f'+4H^{4}(-2q^{3}\nonumber\\-8q^{2}+q-2j-6qj+s+3)f''
+96H^{8}(2q^{2}+3q+j)^{2}f''']\geq0,
\end{eqnarray}
\begin{eqnarray}
\rho-p-48\lambda H^{4}p[qf'+24H^{4}(2q^{2}+3q+j)f'']+48\lambda
H^{4}p[q f'+4H^{4}(-2q^{3}\nonumber\\-8q^{2}+q-2j-6qj+s+3)f''
+96H^{8}(2q^{2}+3q+j)^{2}f''']\geq0,
\end{eqnarray}
\begin{equation}\label{45}
\rho-48\lambda H^{4}p[qf'+24H^{4}(2q^{2}+3q+j)f'']\geq0.
\end{equation}

\subsection{Energy Conditions for specific f(G) models}

~~~In order to exemplify how to use the energy conditions to
constrain the f(G) theories of gravity, below, we study the
realistic models of f(G) gravity, which have been found to reproduce
the current acceleration\cite{15,16}:
\begin{equation}\label{46}
\ f_{1}(G)=\frac{a_{1}G^{n}+b_{1}}{a_{2}G^{n}+b_{2}},
\end{equation}
\begin{equation}\label{47}
\ f_{2}(G)=a_{3}G^{n}(1+b_{3}G^{m}),
\end{equation}
where $a_{1},a_{2},a_{3},b_{1},b_{2},b_{3},n,m$ are all constants.
Since there has been no reliable measurement for the snap parameter
$(s)$ up to now, we only focus on the WEC (\ref{39}) and (\ref{45})
in this particular case.

Since the inequalities are too complicated to find exact analytical
expressions, so we have to take some specific values of the
parameters, such as $a_{1}=b_{1}=-1,a_{2}=2,b_{2}=b$\cite{19}. Also,
when $G\rightarrow\pm\infty$ or $G\rightarrow0^{-}$, the model
(\ref{47}) can be changed in the form $f(G)\sim\alpha
G^{n}$\cite{16}, which means $a_{3}=\alpha, b_{3}=0$. The above two
models can be rewritten as:
\begin{equation}\label{48}
\ f_{1}(G)=-\frac{G^{n}+1}{2G^{n}+b},
\end{equation}
\begin{equation}\label{49}
\ f_{2}(G)=\alpha G^{n}.
\end{equation}

(1) When $L_{m}=-\rho$, we can get the corresponding WEC as follows:
\begin{eqnarray}\label{50}
\rho\{1+48\lambda q H^{4}\frac{G^{n-1}(-nb+2n)}{(2G^{n}+b)^{2}}
+1152\lambda
H^{8}(2q^{2}+3q+j)\nonumber\\\times\frac{G^{n-2}n[2G^{n}(n+1)-(n-1)b](-2+b)}{(2G^{n}+b)^{3}}\}\geq0
~~(for f_{1}(G)),
\end{eqnarray}
\begin{eqnarray}\label{51}
\rho[1+48\lambda q H^{4}\alpha nG^{n-1}+1152\lambda
H^{8}(2q^{2}+3q+j)\alpha n(n-1)G^{n-2}]\geq0 ~~(for f_{2}(G)).
\end{eqnarray}

After a series of simplification, taking some present values
$H_{0}=70.5$\cite{24}, $q_{0}=-0.81\pm0.14$ and
$j_{0}=2.16^{+0.81}_{-0.75}$\cite{25}, and $\lambda=1$, we can
obtain the restrictions on the parameters $n$, $b$ and $\alpha$,
which satisfy the WEC in Eqs.(\ref{50}) and (\ref{51}), respectively
(see Figs.1 and 2). Fig.1 shows that the WEC $\rho^{eff}>0$ can give
the constraints on the parameters $n$ and $b$ in the $f_{1}(G)$
model, i.e., $0\leq b\leq1$ and $-2.3\leq n\leq2.5$, but except
$|n|<0.1$ owing to the non-continuity of $\rho^{eff}>0$. Similarly,
Fig.2 illustrates that only when $\alpha>0$ and $7.5\lesssim
n\lesssim10$, the WEC $\rho^{eff}>0$ can be satisfied in the
$f_{2}(G)$ model.

(2) When $L_{m}=p$, let us consider a perfect fluid composed of
non-relativistic or relativistic particles with constant barotropic
equation of state (EoS) $p=(\gamma-1)\rho$, $0\leq\gamma\leq2$ is a
constant relating to the EoS by $\omega=\gamma-1$\cite{26,27,28},
the corresponding WEC changes into:
\begin{eqnarray}\label{52}
\rho\{1-48\lambda q\omega
H^{4}\frac{G^{n-1}(-nb+2n)}{(2G^{n}+b)^{2}} -1152\lambda
H^{8}\omega(2q^{2}+3q+j)\nonumber\\\times\frac{G^{n-2}n[2G^{n}(n+1)-(n-1)b](-2+b)}{(2G^{n}+b)^{3}}\}\geq0~~(for
f_{1}(G)),
\end{eqnarray}
\begin{eqnarray}\label{53}
\rho[1-48\lambda q\omega H^{4}\alpha nG^{n-1}-1152\lambda
H^{8}\omega(2q^{2}+3q+j)\alpha n(n-1)G^{n-2}]\geq0~~(for f_{2}(G)).
\end{eqnarray}
By the similar discussions to the case of $L_{m}=-\rho$, we can
obtain the restrictions on the parameters $n$, $b$ and $\alpha$,
which respectively satisfy the WEC in Eqs.(\ref{52}) and (\ref{53}),
and are illustrated in Figs.3 and 4 (here taking $\omega=0.5$). From
Fig.3 it is easy to see that the constraints on the parameters $n$
and $b$ for the $f_{1}(G)$ model are $-0.2\leq b\leq1$ and $-2.3\leq
n\leq2.5$, which are nearly the same as the results of the
$L_{m}=-\rho$, but in the Fig.4 $-10\leq \alpha\leq10$ and $-2.7\leq
n\leq-1.3$ for the $f_{2}(G)$ model are quite different from the
results of the $L_{m}=-\rho$.

\vspace{1.0cm}

\begin{figure}[!htb] \vspace{-0.6cm} \hspace{-0.6cm}
\centering
\includegraphics[width=190pt,height=140pt]{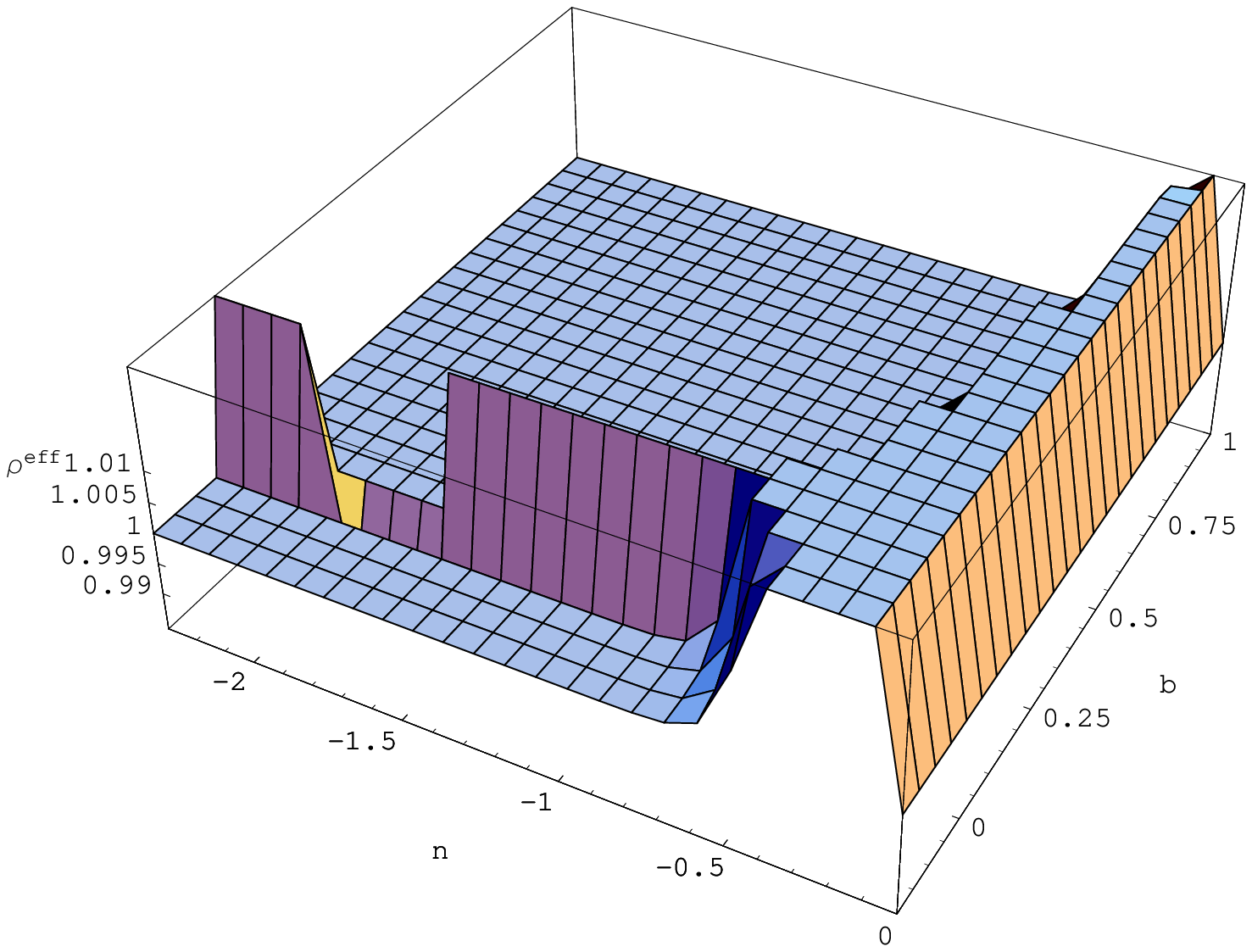} \hspace{-0cm}
\includegraphics[width=190pt,height=140pt]{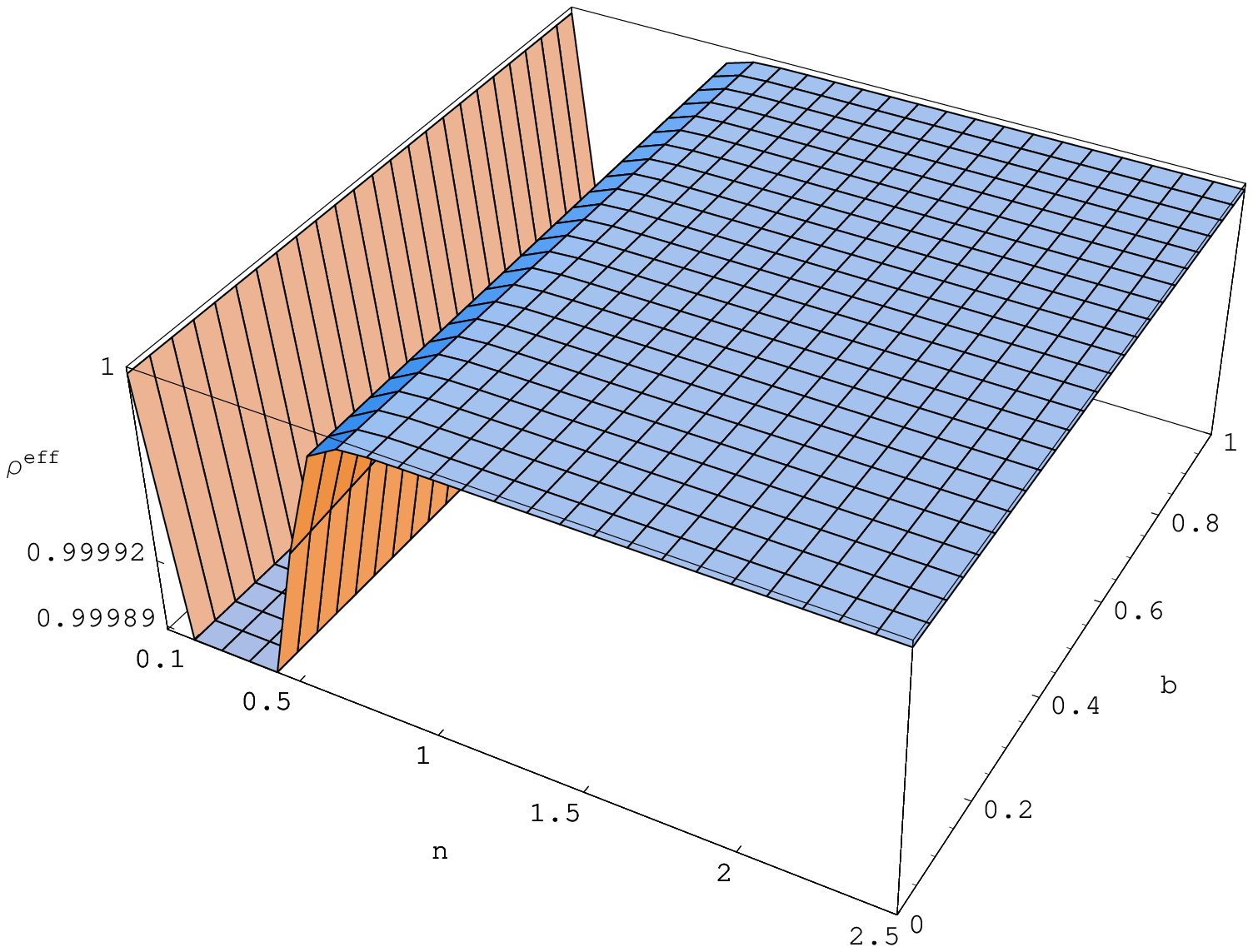} \hspace{-1.2cm}
\\~~(a)~~~~~~~~~~~~~~~~~~~~~~~~~~~~~~(b)~~\\ \caption[Fig.1]{\small{The constraints of WEC ($\rho^{eff}>0$) on
the parameters n and b for the $f_{1}(G)$ model in (\ref{48}) with
$\lambda=1$ and $L_{m}=-\rho$.}} \label{F1}
\end{figure}

\begin{figure}[!htb] \vspace{-0.6cm} \hspace{-0.6cm}
\centering
\includegraphics[width=190pt,height=140pt]{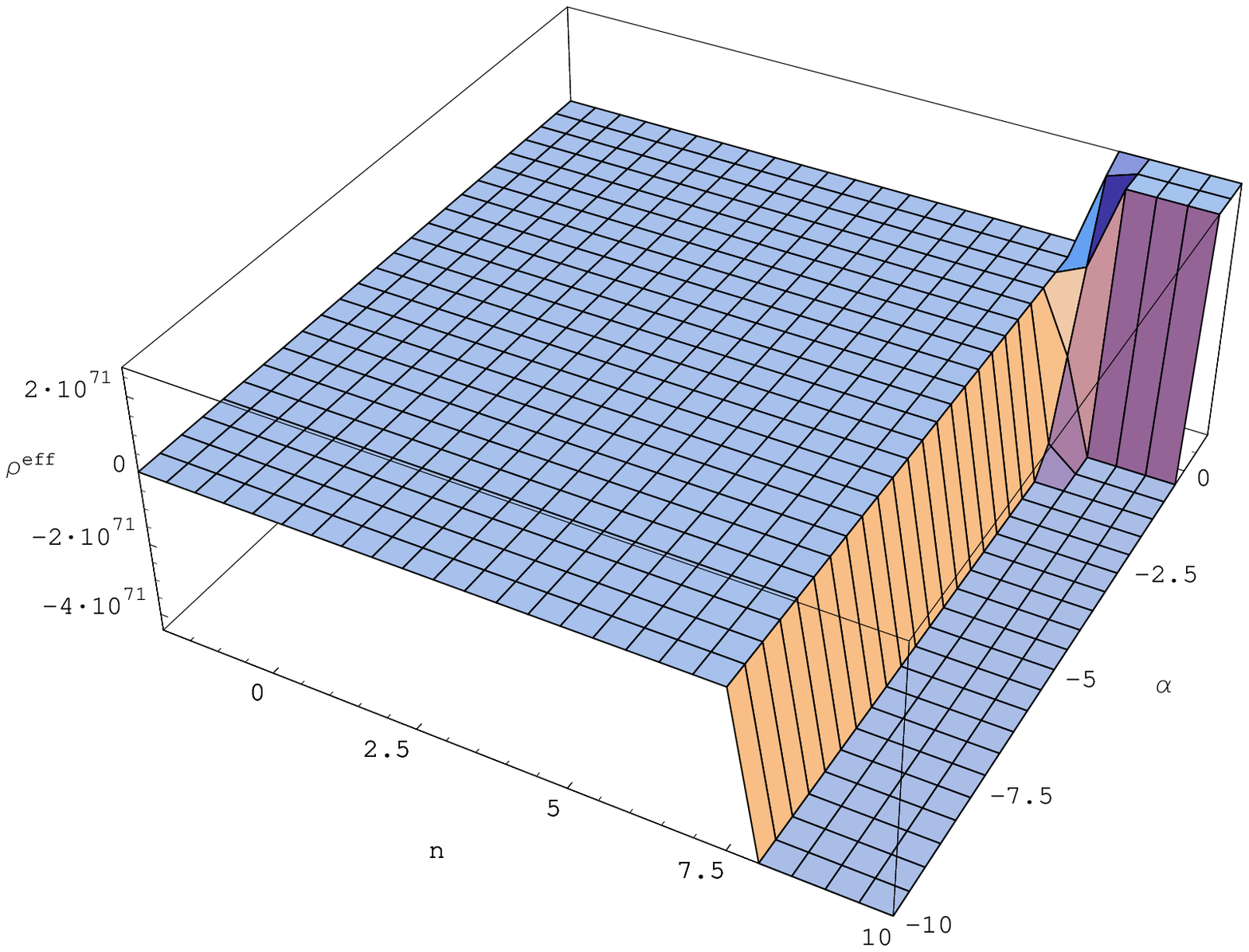} \hspace{-0cm}
\includegraphics[width=190pt,height=140pt]{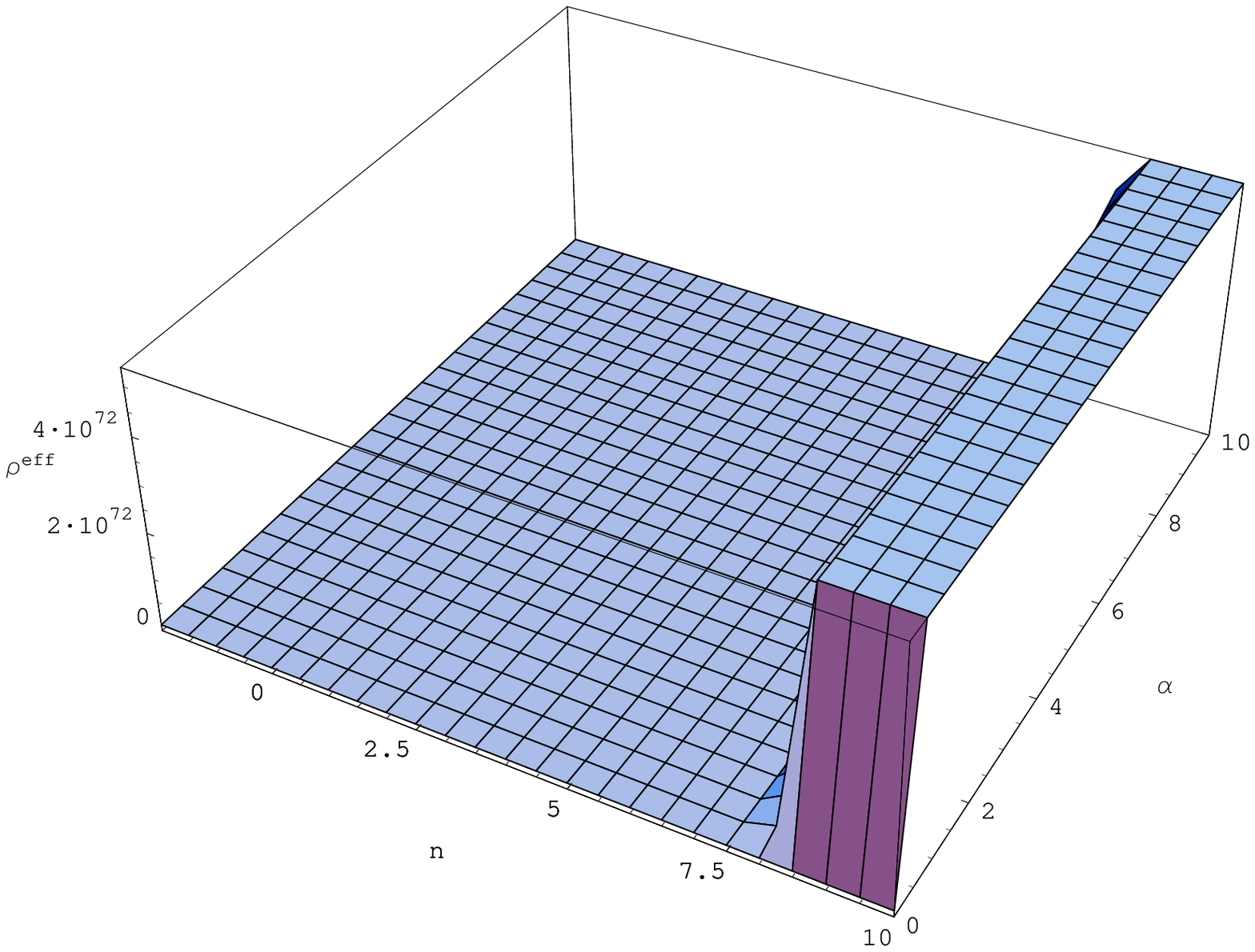} \hspace{-1.2cm}
\\~~(a)~~~~~~~~~~~~~~~~~~~~~~~~~~~~~~(b)~~\\ \caption{\small{The constraints of WEC ($\rho^{eff}>0$) on the
parameters n and $\alpha$ for the $f_{2}(G)$ model in (\ref{49})
with $\lambda=1$ and $L_{m}=-\rho$.}} \label{F2}
\end{figure}

\begin{figure}[!htb] \vspace{-0.6cm} \hspace{-0.6cm}
\centering
\includegraphics[width=190pt,height=140pt]{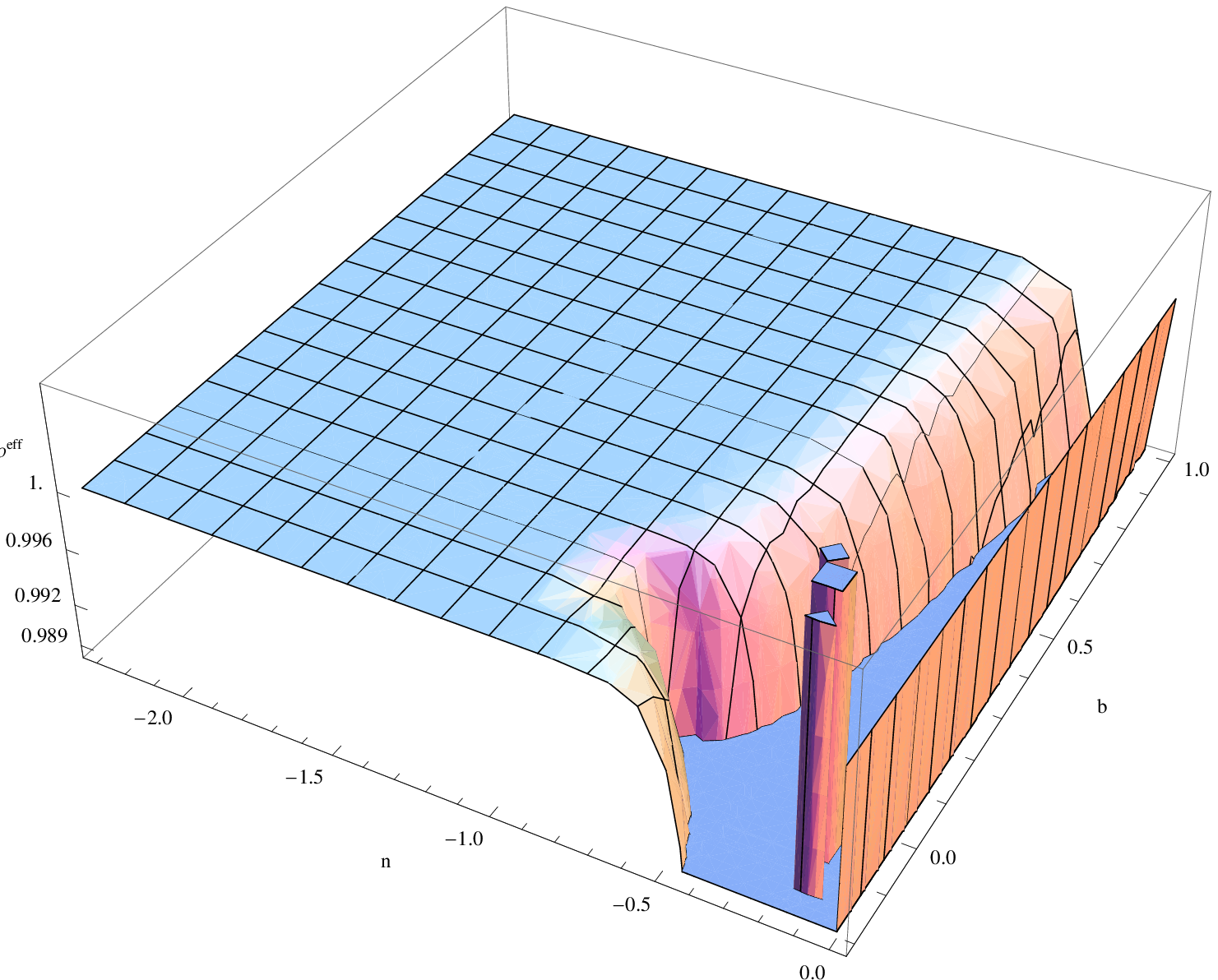} \hspace{-0cm}
\includegraphics[width=190pt,height=140pt]{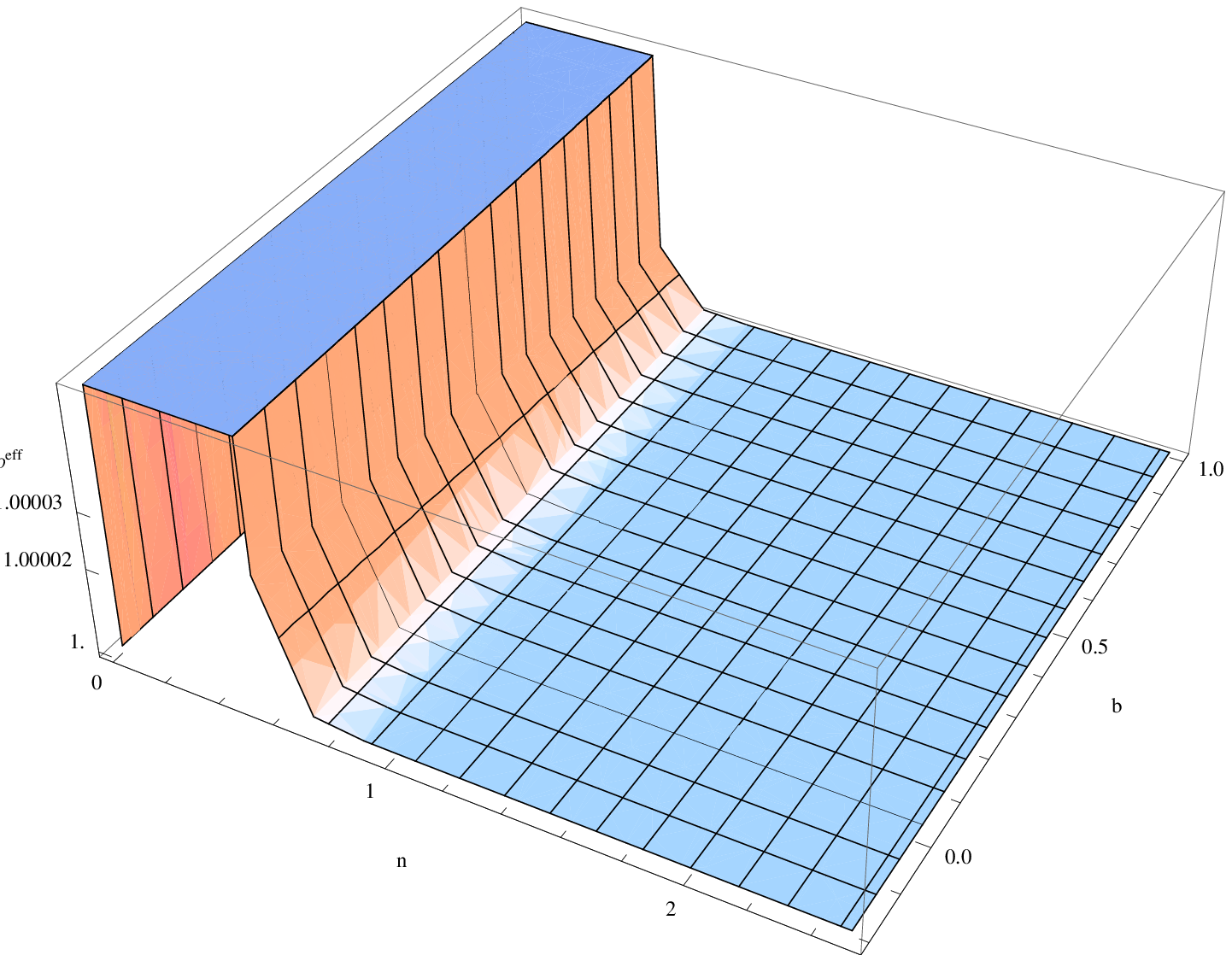} \hspace{-1.2cm}
\\~~(a)~~~~~~~~~~~~~~~~~~~~~~~~~~~~~~(b)~~\\ \caption{\small{The constraints of WEC ($\rho^{eff}>0$) on the
parameters n and $\alpha$ for the $f_{1}(G)$ model in (\ref{48})
with $\lambda=1$ and $L_{m}=p=\omega\rho$ $(\omega=0.5)$.}}
\label{F2}
\end{figure}

\begin{figure}[!htb] \vspace{0.4cm} \hspace{-0.6cm}
\centering
\includegraphics[width=190pt,height=140pt]{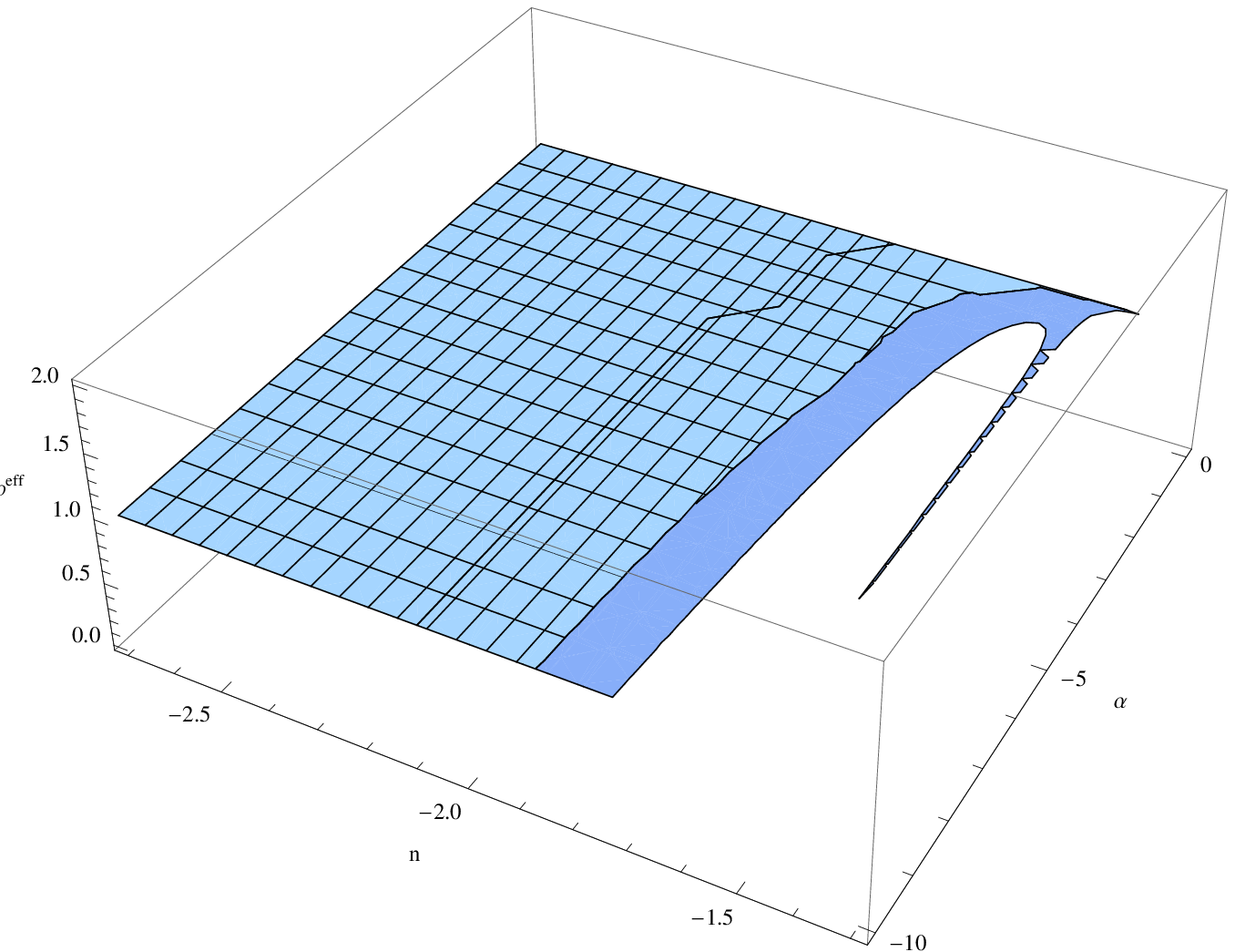} \hspace{-0cm}
\includegraphics[width=190pt,height=140pt]{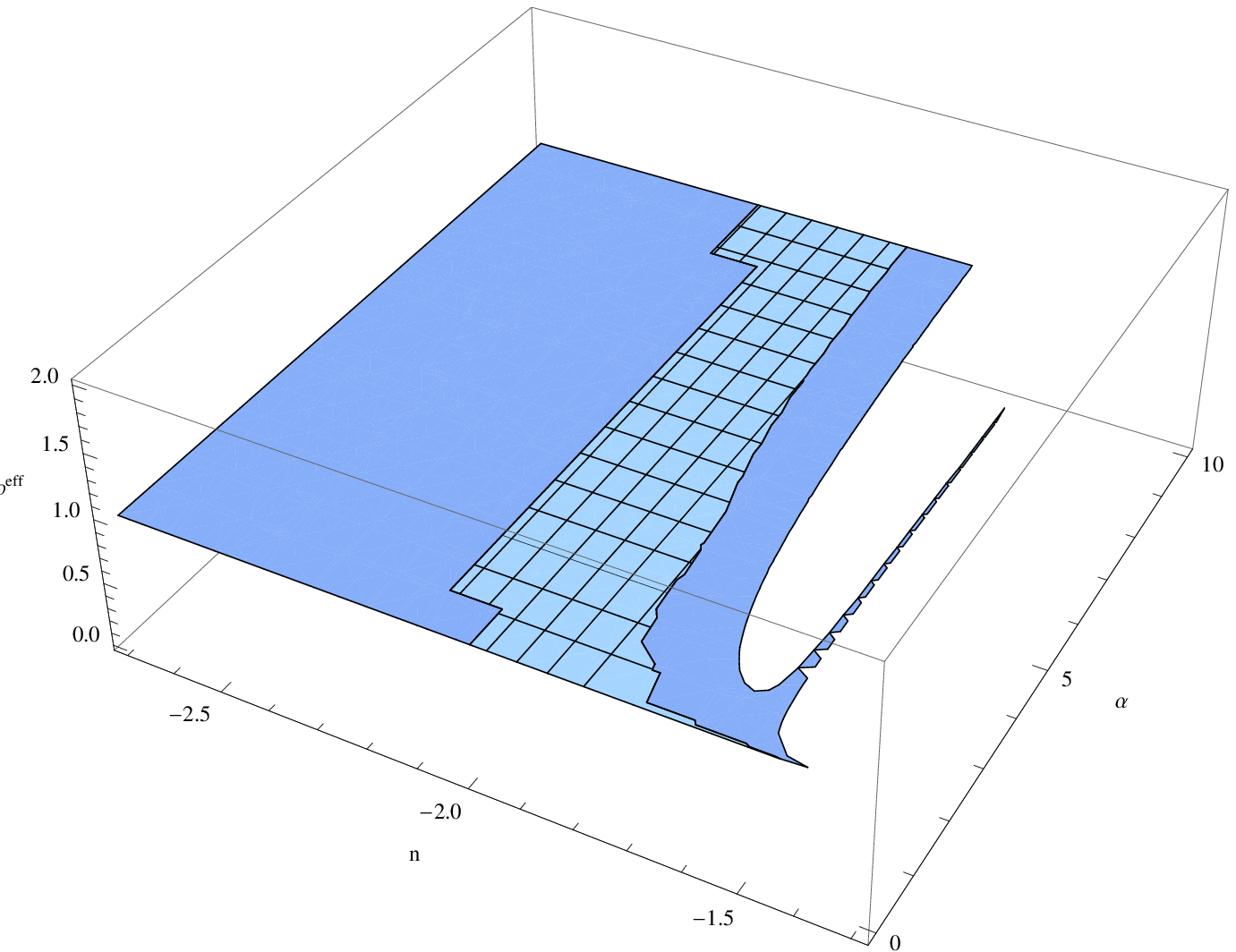} \hspace{-1.2cm}
\\~~(a)~~~~~~~~~~~~~~~~~~~~~~~~~~~~~~(b)~~\\ \caption{\small{The constraints of WEC ($\rho^{eff}>0$) on the
parameters n and $\alpha$ for the $f_{2}(G)$ model in (\ref{49})
with $\lambda=1$ and $L_{m}=p=\omega\rho$ $(\omega=0.5)$.}}
\label{F4}
\end{figure}

\section{Stability criterion at de Sitter point}

~~~Modified gravity must be stable at the classical and quantum
level. There are in principle several kinds of instabilities to
consider, such as Dolgov-Kawasaki criterion in f(R)
gravity\cite{29}. Below, following Ref.\cite{17}, we will focus on
the stability criterion at de Sitter point in the modified f(G)
gravity.

In a flat FLRW background with the metric
\begin{equation}\label{54}
ds^{2}=-dt^{2}+a(t)^{2}dX_{3}^{2},
\end{equation}
where $a(t)$ is the scale factor and $dX^{2}_{3}$ contains the
spacial part of the metric. The 00 component of the field equation
(\ref{3}) gives
\begin{equation}\label{55}
3H^{2}=2\lambda L_{m}Gf_{,G}-48\lambda
L_{m}H^{3}\dot{f_{,G}}+\rho_{m}+\rho_{r}.
\end{equation}

Let us first discuss the stability around the de Sitter point in the
modified f(G) gravity by neglecting the contribution of
pressure-less matter $\rho_{m}$ and radiation $\rho_{r}$. The Hubble
parameter, $H=H_{1}$ (at the de Sitter point), satisfies
\begin{equation}\label{56}
3H_{1}^{2}=2\lambda L_{m}G_{1}f_{,G}(G_{1})-48\lambda
L_{m}H_{1}^{3}\dot{f_{,G}}(G_{1}),
\end{equation}
where $G_{1}=24H_{1}^{4}$, and the relations $\dot{H_{1}}=0$ and
$\dot{G_{1}}=0$ are used. Considering a linear perturbation $\delta
H_{1}$ about the de Sitter point, Eq.(\ref{55}) gives
\begin{equation}\label{57}
2\lambda L_{m}(24H_{1}^{3}f_{,GG}\delta \dot{G_{1}}-f_{,G}\delta
G_{1})=\delta H_{1}(48\lambda L_{m}H_{1}^{4}f_{,GG}-6H_{1}).
\end{equation}
Substituting the relations $\delta G(H_{1})=24(4H_{1}^{3}\delta
H_{1}+H_{1}^{2}\delta \dot{H_{1}})$ and $\delta
\dot{G}(H_{1})=24H_{1}^{2}(\delta \ddot{H_{1}}+4H_{1}\delta
\dot{H_{1}})$ into Eq.(\ref{57}), we obtain
\begin{equation}\label{58}
\delta
\ddot{H_{1}}+\frac{96H_{1}^{4}f_{,GG}-f_{,G}}{24H_{1}^{3}f_{,GG}}\delta
\dot{H_{1}}+(\frac{1}{192\lambda
L_{m}H_{1}^{4}f_{,GG}}-\frac{1}{24H_{1}}-\frac{f_{,G}}{6H_{1}^{2}f_{,GG}})\delta
H_{1}=0.
\end{equation}

It follows that the effective mass squared is $(\frac{1}{192\lambda
L_{m}H_{1}^{4}f_{,GG}}-\frac{1}{24H_{1}}-\frac{f_{,G}}{6H_{1}^{2}f_{,GG}})$,
which must be non-negative for stability. Therefore, we can obtain
\begin{equation}\label{59}
\frac{1-8\lambda H_{1}^{3}f_{,GG}-32\lambda
L_{m}H_{1}^{2}f_{,G}}{192\lambda L_{m}H_{1}^{4}f_{,GG}}>0,
\end{equation}
which is just the stability criterion at the de Sitter point. It
follows that if the exact value of $H_{1}$ and a suitable form of
$L_{m}$ can be given in the modified f(G) gravity models, then the
constraints on the parameters in the specific model can be obtained.

\section{The conditions for late-time cosmic accelerated expansion in the modified f(G) gravity}

~~~It is known that late-time cosmic accelerated expansion occurs
under the conditions of either a power-law expansion or the equation
of state of matter less than $-\frac{1}{3}$. To exemplify how to use
these conditions to realize the phase of accelerating expansion in
the modified f(G) gravity, now we concentrate on the model
$f_{1}(G)$ in (\ref{48}). Thus, by means of the action (\ref{2}) and
the energy density $L_{m}$ of perfect fluid\cite{30,31}, i.e.,
\begin{equation}\label{60}
L_{m}=-\rho=-\rho_{0}a^{-3(1+\omega)},
\end{equation}
where $\omega$ is the equation of state of perfect fluid and is
taken to be a constant, the field equation (\ref{55}) becomes
\begin{equation}\label{61}
\ 3H^{2}=-\rho_{0}a^{-3(1+\omega)}[1-48\lambda
H^{2}(H^{2}+\dot{H})f'-1152H^{4}(2\dot{H}^{2}+H\ddot{H}+4H^{2}\dot{H})f''].
\end{equation}
Note that the relation $R=6(2H^{2}+\dot{H})$ is used in the
derivation of the Eq.(\ref{61}).

If assuming the solution of (\ref{61}) is $a=a_{0}t^{r}$\cite{13},
then we have $H=\frac{r}{t}$, $\dot{H}=-\frac{r}{t^{2}}$,
$\ddot{H}=\frac{2r}{t^{3}}$. Substituting all these relations into
(\ref{61}), we can get the following equation:
\begin{eqnarray}
3r^{2}&=&\rho_{0}a_{0}^{-3(1+\omega)}t^{-3r(1+\omega)+2}\frac{1}{(r-1)[b+2^{3n+1}\times3^{n}
\frac{(r-1)^{n}r^{3n}}{t^{4n}}]^{3}} \times \{b^{3}(r-1)\nonumber
\\& &+8\frac{(r-1)^{2n}r^{6n}}{t^{8n}}[-13824^{n}\times\frac{(r-1)^{n}r^{3n}}{t^{4n}}
+13824^{n}\times \frac{r(r-1)^{n}r^{3n}}{t^{4n}}
+5576^{n}n\lambda \nonumber\\
& & +4^{3n+1}\times9^{n}n^{2}\lambda-576^{n}nr\lambda]
+2^{3n+1}\times3^{n}b^{2}\frac{(r-1)^{n}r^{3n}}{t^{4n}}
[-3-5n\lambda+4n^{2}\lambda \nonumber\\&&+r(3+n\lambda)]
+4b\frac{(r-1)^{n}r^{3n}}{t^{4n}}\{-3^{2n+1}\times64^{n}\frac{(r-1)^{n}r^{3n}}{t^{4n}}
+5n\lambda[24^{n}\nonumber\\&
&-576^{n}\frac{(r-1)^{n}r^{3n}}{t^{4n}}]
-4n^{2}\lambda[24^{n}+2^{6n+1}\times9^{n}\frac{(r-1)^{n}r^{3n}}{t^{4n}}-576^{n}
\frac{(r-1)^{n}r^{3n}}{t^{4n}}]\nonumber\\&&+r[3^{2n+1}\times64^{n}\frac{(r-1)^{n}r^{3n}}{t^{4n}}
-24^{n}n\lambda+576^{n}n\lambda\frac{(r-1)^{n}r^{3n}}{t^{4n}}]\}\}.
\end{eqnarray}

We find six kinds of possible relationships among r, $\omega$ and n,
namely, $r=\frac{2(1-2n)}{3(1+\omega)}$,
$r=\frac{2(1-4n)}{3(1+\omega)}$, $r=\frac{2(1-6n)}{3(1+\omega)}$,
$r=\frac{2(1-8n)}{3(1+\omega)}$, $r=\frac{2(1-10n)}{3(1+\omega)}$
and $r=\frac{2(1-12n)}{3(1+\omega)}$. Under the condition of a
power-law expansion (i.e., $r>1$), the corresponding regions of
$\omega$ are $\omega < -\frac{1}{3}-\frac{4}{3}n$ for
$r=\frac{2(1-2n)}{3(1+\omega)}$, $\omega <
-\frac{1}{3}-\frac{8}{3}n$ for $r=\frac{2(1-4n)}{3(1+\omega)}$,
$\omega < -\frac{1}{3}-4n$ for $r=\frac{2(1-6n)}{3(1+\omega)}$,
$\omega < -\frac{1}{3}-\frac{16}{3}n$ for
$r=\frac{2(1-8n)}{3(1+\omega)}$, $\omega <
-\frac{1}{3}-\frac{20}{3}n$ for $r=\frac{2(1-10n)}{3(1+\omega)}$ and
$\omega < -\frac{1}{3}-8n$ for $r=\frac{2(1-12n)}{3(1+\omega)}$,
respectively. Furthermore, by considering the equation of state of
matter less than $-\frac{1}{3}$ (i.e., $\omega < -\frac{1}{3}$), the
relationship among $r$, $\omega$ and $n$, condition and candidate
for late-time cosmic accelerated expansion are shown in Table 1. The
candidate for late-time cosmic accelerated expansion can be either
the effective quintessence ($-1 < \omega< -\frac{1}{3}$) or the
effective phantom ($\omega < -1$).

\begin{table}[!htb]
\centering \hspace{0.1cm}\small{\begin{tabular}{|c|c|c|c|} \hline
Relationship & Condition($r > 1, \omega < -\frac{1}{3}$) & The ~effective~ quintessence & The~ effective~ phantom \\
\hline $r=\frac{2(1-2n)}{3(1+\omega)}$ & $n > 0$ and $n\neq\frac{1}{2}$ & $0 < n < \frac{1}{2}$ & $n > \frac{1}{2}$\\
\hline $r=\frac{2(1-4n)}{3(1+\omega)}$ & $n > 0$ and $n\neq\frac{1}{4}$ & $0 < n < \frac{1}{4}$ & $n > \frac{1}{4}$\\
\hline $r=\frac{2(1-6n)}{3(1+\omega)}$ & $n > 0$ and $n\neq\frac{1}{6}$ & $0 < n < \frac{1}{6}$ & $n > \frac{1}{6}$\\
\hline $r=\frac{2(1-8n)}{3(1+\omega)}$ & $n > 0$ and $n\neq\frac{1}{8}$ & $0 < n < \frac{1}{8}$ & $n > \frac{1}{8}$\\
\hline $r=\frac{2(1-10n)}{3(1+\omega)}$ & $n > 0$ and $n\neq\frac{1}{10}$ & $0 < n < \frac{1}{10}$ & $n > \frac{1}{10}$\\
\hline $r=\frac{2(1-12n)}{3(1+\omega)}$ & $n > 0$ and $n\neq\frac{1}{12}$ & $0 < n < \frac{1}{12}$ & $n > \frac{1}{12}$\\
\hline
\end{tabular}}
\caption{\small{The relationship among $r$, $\omega$ and $n$,
condition and candidate for late-time cosmic accelerated expansion
in case $f(G)=-\frac{G^{n}+1}{2G^{n}+b}$.}} \label{T1}
\end{table}

From the above discussions, it is easy to see that the results in
the model are interesting. Compared with the f(R) models, f(G)
models are even more complicated. Since the Hubble parameter can be
expressed as $H=\frac{r}{t}$, the GB term turns into
$G=\frac{24r^{3}(r-1)}{t^{4}}$. If $0 < r < 1$, the early universe
is in deceleration phase, which corresponds to the matter dominated
phase with $r=\frac{2}{3}$, and if $r > 1$, the late universe is in
acceleration phase.

Note that for the case of $L_{m}=p=\omega\rho$ we can make similar
discussions to ones in the case of $L_{m}=-\rho$, and obtain the
same results as the ones shown in Table 1 due to the constant
$\omega$.

\section{Conclusions and discussions}

~~~In the present paper we have considered a modified f(G) gravity
model with coupling between matter and geometry, described by the
product of the Lagrange density of the matter and an arbitrary
function of the Gauss-Bonnet term. The proposed action represents
the general extension of the standard Hilbert action for the
gravitational field, $S=\int d^4x\sqrt{-g}\{\frac{R}{2}+[1+\lambda
f(G)]L_m\}$. The field equations and the equations of motion
corresponding to this model show the non-conservation of the
energy-momentum tensor, the presence of an extra-force acting on
test particles and the non-geodesic motion. Moreover, in the
modified f(G) gravity we have derived the energy conditions (SEC,
NEC, DEC, WEC) when we consider $L_{m}=-\rho$ and $L_{m}=p$,
respectively. For the SEC and the NEC, the Raychaudhuri equation,
which is the physical origin of them, has been used. From the
derivation, we found equivalent results can be obtained by taking
the transformations $\rho\rightarrow\rho^{eff}$ and $p\rightarrow
p^{eff}$ into $\rho+3p\geq0$ and $\rho+p\geq0$. By means of these
transformations, the DEC and WEC in the modified f(G) gravity have
been also obtained. In order to exemplify how to use these energy
conditions to constrain the modified f(G) gravity models, we have
considered two specific models of f(G) gravity, i.e., $f_{1}(G)$ and
$f_{2}(G)$ and given the corresponding constraints on the parameters
in the $f_{1}(G)$ and $f_{2}(G)$ models. Since there has been no
reliable measurement for the snap parameter $(s)$ up to now, we only
focus on the WEC in this particular case. By analysis on Figs.1 and
2 we have given the constraints on the parameters in the $f_{1}(G)$
 and $f_{2}(G)$ models satisfying the weak energy conditions when $L_{m}=-\rho$.
By the similar discussions to the case of $L_{m}=-\rho$, when
$L_{m}=p$ the restrictions on the parameters $n$, $b$ and $\alpha$
have been also illustrated in Figs.3 and 4, from which we have found
that in the two different forms of $L_{m}$ the constraints on the
parameters for the $f_{1}(G)$ model are nearly the same, but quite
different for the $f_{2}(G)$ model. Furthermore, we have derived the
stability criterion at the de Sitter point for the modified f(G)
gravity models, which means that the modified f(G) gravity models
may be stable. In addition, we have researched the conditions for
late-time cosmic accelerated expansion in the modified f(G) gravity.
Concretely, for the two different forms of $L_{m}$, the relationship
among r, $\omega$ and n have been respectively given in the model
$f(G)=-\frac{G^{n}+1}{2G^{n}+b}$, and by using the conditions of
power-law accelerated expansion and the equation of state of matter
less than $-\frac{1}{3}$, the constraints on the parameter n have
been obtained, which are exactly the same in the two different forms
of $L_{m}$. The candidate for late-time cosmic accelerated expansion
would be either the effective quintessence ($-1 < \omega<
-\frac{1}{3}$) or the effective phantom ($\omega < -1$), which could
be determined by choosing n properly. Of course, other forms of f(G)
gravity models with curvature-matter coupling will be considered in
our following investigations.

 \textbf{\ Acknowledgments}
 The research work is supported by   the National Natural Science Foundation of
China (11147150), the Natural Science Foundation of Education
Department of Liaoning Province (L2011189), the Natural Science
Foundation of Liaoning Province, China (Grant No.20102124)
 the NSFC (11175077)  of P.R. China.

\end{document}